\RequirePackage{fix-cm}
\documentclass{svjour3}                     
\smartqed  
\usepackage{graphicx}
\journalname{Computational Astrophysics and Cosmology}
\begin{document}

\title{Riemann Solvers and Alfven Waves in Black Hole
Magnetospheres
}


\author{Brian Punsly\and Dinshaw Balsara\and
Jinho Kim\and Sudip Garain 
}


\institute{B.Punsly \at
              1415 Granvia Altamira, Palos Verdes Estates CA, USA
90274 \\
             ICRANet, Piazza della Repubblica 10 Pescara 65100, Italy\\
              \email{brian.punsly@cox.net}           
           \and
           D. Balsara\and
J. Kim\and S. Garain \at
              Physics Department,
University of Notre Dame du Lac, 225 Nieuwland Science Hall, Notre
Dame, IN 46556, USA }


\maketitle

\begin{abstract}
In the magnetosphere of a rotating black hole, an inner Alfven
critical surface (IACS) must be crossed by inflowing plasma. Inside
the IACS, Alfven waves are inward directed toward the black hole.
The majority of the proper volume of the active region of spacetime
(the ergosphere) is inside of the IACS. The charge and the totally
transverse momentum flux (the momentum flux transverse to both the
wave normal and the unperturbed magnetic field) are both determined
exclusively by the Alfven polarization. Thus, it is important for
numerical simulations of black hole magnetospheres to minimize the
dissipation of Alfven waves. Elements of the dissipated wave emerge
in adjacent cells regardless of the IACS, there is no mechanism to
prevent Alfvenic information from crossing outward. Thus, numerical
dissipation can affect how simulated magnetospheres attain the
substantial Goldreich-Julian charge density associated with the
rotating magnetic field. In order to help minimize dissipation of
Alfven waves in relativistic numerical simulations we have
formulated a one-dimensional Riemann solver, called HLLI, which
incorporates the Alfven discontinuity and the contact discontinuity.
We have also formulated a multidimensional Riemann solver, called
MuSIC, that enables low dissipation propagation of Alfven waves in
multiple dimensions. The importance of higher order schemes in
lowering the numerical dissipation of Alfven waves is also
catalogued. 
\end{abstract}

\section{Introduction}
\label{intro} In this century, there has been great progress in 3-D
magnetohydrodynamic (MHD) simulations of black hole magnetospheres
\cite{dev03,kom04,kom05,haw06,fra07,mck09,mck12}. To varying
degrees, each of these simulations require knowledge of the 1-D
characteristics of the MHD system in order to time evolve the
magnetosphere. Specifically, the polarization properties of the
waves determine the changes in the fields that can be propagated at
the appropriate speed along a particular characteristic direction.
In single fluid ideal MHD there are three plasma modes in the
system, the fast mode, the Alfven (or intermediate) mode and the
slow mode. Black hole magnetospheres have the property that all
plasma must pass progressively through the slow, Alfven and fast
critical surfaces before reaching the event horizon \cite{pun08}. As
each critical surface is crossed, the unique information associated
with each wave mode is unable to be communicated upstream to an
outgoing wind or jet. The event horizon wind system has no boundary
conditions at its terminus, there are asymptotic infinities both at
the event horizon and at large radial coordinate \cite{pun08}. There
are only lateral boundary
 conditions imposed by accreting gas. Thus, the wind system itself
 and the lateral boundary conditions determine 3-D single fluid
 perfect MHD wind solutions. Furthermore, due to the paired wind
 nature of the event horizon wind system (an ingoing accretion inner wind and the outgoing outer wind or jet), plasma is always drained
 off of the field lines and auxiliary physics (mass floors) must be
 injected by hand in order to keep numerical simulations from crashing
 at low density. Mass floors are a source of MHD waves and are generally
 chosen to enhance dissipation. Consequently, there are many unique aspects to the application
 of MHD that can influence the final steady state of the wind
 system. Describing the evolution of the event horizon magnetosphere
 with single fluid MHD is wrought with non trivial subtleties.

 \par These subtleties relate to the numerical determination of the field line
 angular velocity as viewed from asymptotic infinity, $\Omega_{F}$.
 This is of primary interest since the Poynting flux of the wind
 scales as $\Omega_{F}^{2}$ \cite{pun08}. First, contrary to previous claims of
 early simulations, newer simulations indicate that $\Omega_{F}$ can
 be altered significantly by the auxiliary method of injecting
 plasma \cite{mck12,bes13}. We consider the
unique role of the oblique Alfven wave in this process. A unique component of the momentum flux is
propagated along the Alfven characteristics and this momentum flux
is a component of the MHD equations written in conservative form. It
is also the only isolated discontinuity that propagates a physical
charge. A black hole magnetospheres that support an outgoing
relativistic jet, rotate and have a Goldreich-Julian charge density.
The Alfven critical surface for the inflow (IACS) is quite far from
the event horizon. For rapidly rotating black holes and the most
recent $\Omega_{F}$ values from numerical simulations, the proper
distance is 2 to 3 black hole radii from the event horizon. For
proper evolution of the magnetic field rotation rate and the induced
charge density, one must be able to simulate the role of Alfven
waves with high fidelity both globally and inside of the IACS.
Thusly motivated, we discuss in this article new numerical methods
that are designed to accurately characterize the Alfven wave
numerically in the rarefied environment of black hole magnetosphere.

\par An accurate depiction of the time evolution
of a black hole magnetosphere and the global considerations germane
to the IACS are intimately related to minimizing the numerical
dissipation of Alfven waves. The IACS is a one-way surface as far as
t he propagation of Alfven waves is concerned. In other words, at
the IACS and within it, all Alfven waves should propagate inwards
and only inwards. The propagation of waves in a higher order Godunov
code is modulated by the Riemann solver. It is, therefore desirable
if the Riemann solver can mimic this one-way propagation property
for Alfven waves. Alas, whether a Riemann solver does so or not,
depends crucially on the design of the Riemann solver. Some Riemann
solvers which retain the substructure associated with Alfven waves
within the Riemann fan, can indeed represent such a one-way
propagation of Alfven waves. Other Riemann solvers, like the HLL
Riemann solver, do not retain the substructure associated with the
Alfven waves. At extraordinarily high resolutions, any well-designed
code will of course minimize this dissipation. However, the present
generation of simulations have all been done with low or modest
resolutions. Furthermore, they have mostly used the HLL Riemann
solver which, we argue, applies a maximal, and deleterious,
dissipation to Alfven waves. To appreciate this point, realize that
the HLL Riemann solver is based on a wave model that has just two
extremal waves. These two extremal waves determine the ends of the
Riemann fan in one-dimension. The speed of these extremal waves is
usually set to the extremal signal speeds in the physical problem.
For a relativistic MHD (RMHD) simulation of highly magnetized event
horizon magnetospheres, these extremal signal speeds are usually set
to approximately the speed of light propagating in either direction
at a zone interface where the Riemann solver is applied. The HLL
Riemann solver does not incorporate any further sub-structure
associated with intermediate waves. Consequently, the HLL Riemann
solver maximizes the dissipation of Alfven waves even near the IACS.
This is the very location where the dissipation of these waves has
to be minimized. Introducing the intermediate waves in the Riemann
fan reduces the dissipation, but that effect is not incorporated in
the HLL Riemann solver.

\par There is another issue that increases the
dissipation of Alfven waves, in more than one dimension. Riemann
solvers applied to black hole magnetospheres  have been treated as
1-D in each direction. However, a true multi-dimensional Riemann
solver has a strongly interacting region in which the numerical
fluxes in orthogonal directions become intertwined \cite{bal12}.
Careful treatment of the strongly-interacting region results in far
less dissipation (as we show for RMHD in Section 4). The increase in
computational complexity associated with a multidimensional Riemann
solver is handily offset by larger timesteps and greater code
robustness. The accuracy of the numerical depiction of the role of
the Alfven waves near a black hole during jet production is
facilitated by a true multidimensional scheme that incorporates the
strongly interacting region.
\par Our ultimate goal is to understand the detailed time evolution
of black hole magnetospheres. This is subtle because one must try to
understand in each time step how the mass floor is affecting the
time evolution. Thus, the transient structure is essential to
monitor in order to see how transients associated with plasma
injection are altering the time evolution of the system. This
requires an inherently very stable numerical scheme
(``well-balanced", which we describe below). Furthermore, we note
that the simulations of \cite{kom04,kom05} did tend to a steady
state and this made it possible to carry out convergence testing for
those simulations. However, several simulations seemed to never
reach a steady state and, therefore, one cannot carry out
convergence studies for them. In particular, an unexpected finding
of \cite{kro05} was that the event horizon magnetospheres in
numerical simulations are very unsteady and appear to be more like a
cauldron of strong MHD waves rather than a force-free structure,
``For example, although the funnel region is magnetically-dominated,
it is not in general in a state of force-free equilibrium. Indeed,
the very large fluctuations that continually occur in the outflow
show that it is never in any state of equilibrium, force-free or
otherwise." This also appears to be the case in the simulations of
\cite{mck12} based on the supporting online movies. This suggests
that wind formation in event horizon magnetospheres might be subject
to large numerically induced transients which would mask the kinds
of effects that we are looking for. Alternatively, if these large
transients are integral to jet formation it indicates a dynamic in
which strong waves from the lateral boundaries created by the
accretion flow scatter off the event horizon magnetosphere producing
strong gusts of energy in a jetted outflow.

\par It is important to separate these potential
physical effects from numerical effects. However, schemes that are
based on higher order Godunov methods, especially those that are
based on the HLL Riemann solver, are notorious for not achieving
steady state even when the physical problem admits such a steady
state! This was first observed when higher order Godunov schemes
with Riemann solvers were first applied to metreological simulations
(which simulate wind flow in the earth's atmosphere) or to the
shallow water equations (which simulate lake and ocean circulation);
\cite{par06}. Unless the numerical scheme has a special property
called well-balancing, it usually does not find a stationary state
even when one exists. Instead, the simulation generates "numerical
storms" -- high velocity flows that are entirely a numerical
artefact. The issue of well-balancing has recently become more
compelling within the context of astrophysical flows with the work
of Kappeli and Mishra (2014, 2016). Within the context of Type IIb
supernova simulations, it has been found that the proto-neutron
stars refuse to reach steady state if the scheme is not
well-balanced, i.e., the numerical method has to be specially
engineered to that it can find a steady-state proto-neutron star
solution if one exists. Kappeli and Mishra have also explored
time-dependent simulations that are not in steady state. Their very
interesting result is that even for simulations that have no reason
to be in steady state, the inclusion of well-balancing provides a
substantial improvement in the accuracy of the simulation. This
result has bearing on the black hole magnetospheres problem because
it suggests that even when the simulations are far from steady state
they might indeed be helped by well-balancing.
\par In order to verify that one has a well-balanced scheme, one has to know what the steady
states of the system are and make sure that the simulated system is
driven towards that steady state. This does not mean that every
system must reach a steady state; it only means that when such a
steady state exists, the numerical code is equipped to find it. For
a scheme to be truly well-balanced, it needs to have two attributes.
First, the reconstruction procedure has to be modified so that any
potential steady state solution is subtracted off from the
reconstructed solution. Second, the scheme must use a Riemann solver
that can capture intermediate waves -- specifically the contact
discontinuity in classical Euler flow. We admit, that in certain
circumstances it may not be possible to identify the steady state
solution that has to be subtracted off. In such situations, the
numerical scheme should at least be well-balanced up to second
order. In practical terms, this means that a modest resolution
simulation will not find the steady state solution to machine
accuracy. However, it will nevertheless find the steady state
solution with accuracies that are proportional to the size of the
mesh squared. Being well-balanced up to second order is a weaker
notion of well-balancing compared to being truly well-balanced. In
order for a scheme to be well-balanced up to second order, it is
imperative that the Riemann solver should at least capture
stationary contact discontinuities in a self-gravitating situation
involving Euler flow. By extension, any MHD code that is capable of
capturing stationary equilibria up to second order should at least
be based on a Riemann solver that captures contact discontinuities
and Alfven waves. The discussion in this paragraph has made it clear
that in order to disentangle a flow that is chaotic because of the
physics of the situation from a flow that is chaotic because of
simple numerical effects, one must pay attention to the form of the
Riemann solver.

\par In summary, for the purposes of exploring the time evolution of
event horizon magnetospheres, we require a well-balanced scheme to
second order that will eliminate recurrent large transients. This is
facilitated by preserving the Alfven wave as discussed above, but
also requires that the inclusion of the contact discontinuity in the
Riemann fan. Furthermore, without the contact discontinuity, the
huge density gradient between accretion flow and the evacuated
funnel of the event horizon magnetosphere means that plasma flows
into the funnel by numerical diffusion. This is far from ideal if we
want to explore the role of mass injection on the final solutions.
To improve this situation, one desires a Riemann solver that at
least minimizes the dissipation at contact discontinuities. (For
more helpful information on numerical schemes and Riemann solvers of
relevance to astrophysics, please visit
http://www.nd.edu/~dbalsara/Numerical-PDE-Course.)

\par Designing low-dissipation Riemann solvers for RMHD is a challenging enterprise.
An exact RMHD Riemann solver exists, \cite{gia06,gia07}, but it is
not practicable for use in numerical codes. HLLC Riemann solvers for
RMHD exist \cite{mb06,hj07,kb14}. They enable a stationary contact
discontinuity to be captured on a mesh. However, they dissipate
stationary Alfven waves just like an HLL Riemann solver. HLLD
Riemann solvers for RMHD, \cite{mig09}, do exist, which enable
stationary contact discontinuities as well as stationary Alfvenic
discontinuities to be captured on a computational mesh.
Unfortunately, the method is iterative, which makes it
computationally very expensive. Furthermore, when an iteration fails
to converge, the method becomes brittle. With the emergence of the
HLLI Riemann solver, \cite{dum16}, it has become possible to capture
stationary contact discontinuities as well as stationary Alfvenic
discontinuities using a Riemann solver that is non-iterative and
computationally inexpensive. The first goal of this paper is to
document this capability in the astrophysical literature.

\par RMHD simulations also have to maintain the divergence-free structure of the magnetic field.
This necessitates the use of a Yee-type mesh where the magnetic
field components are specified at the faces of the mesh and the
electric fields are to be evaluated at the edges of the mesh. It was
claimed by \cite{gs05,gs08,bs11,ws15} that stabilizing the evolution
of the magnetic field requires that one should always double the
dissipation in the electric field at every timestep. Unfortunately,
that approach has been used in the RMHD literature with the result
that the already excessive dissipation of the HLL Riemann solver is
increased even further in simulations. Such explorations ignore
recent advances in multidimensional Riemann solver technology
\cite{bal10,bal12,bal14,bal15,bda14,bd15,bvg16}. In a recent paper,
\cite{bk16}, showed that an exact analogue of the HLLI Riemann
solver in multidimensions can be designed for RMHD. Their work is
based on the original paper by Balsara et al. (2016b). Such
multidimensional Riemann solvers go under the name of MuSIC Riemann
solvers. Here the MuSIC acronym stands for Multidimensional,
Self-similar strongly-Interacting riemann solver that is based on
Consistency with the conservation law. By introducing substructure
associated with the multidimensional propagation of Alfven waves,
the MuSIC Riemann solver reduces the dissipation of Alfven waves
that propagate at any angle with respect to the mesh. The second
goal of this paper is to catalogue the advantages of the MuSIC
Riemann solver in reducing the dissipation involved in the
multidimensional propagation of Alfven waves in RMHD.

\par In this article, we claim that modern 5-wave Riemann solvers can now be
implemented that can allow a systematic assessment of the issues
related to determination of $\Omega_{F}$. This would require the
formulation and simulation of simple magnetospheres, lateral
boundary conditions and plasma injection mechanisms. Proper time
evolution of the 3-D magnetosphere requires two main aspects of the
solver, low dissipation and well balancing. We have motivated both
these issues in the previous paragraphs. We demonstrate that the new
Riemann solvers described here are capable of delivering on these
goals in the following.

\par Riemann solvers offer one way of reducing numerical dissipation
and Riemann solvers that preserve essential features of the flow are
certainly central to many aspects of jet simulation. Recent advances
in higher order schemes has made it possible to go beyond the
traditional second order Godunov scheme that is commonplace in
computational astrophysics. The third goal of this paper is to show
that higher order schemes for RMHD do exist, \cite{bk16,del07,zd16},
which make it possible to go beyond second order of accuracy. We
show that the combination of higher order schemes and appropriate
Riemann solvers can go a long way towards enabling almost
dissipation-free propagation of Alfven waves.

\par The paper is organized as follows. In Section 2, we discuss
dissipation inside the IACS in Riemann solvers in general terms. It
is shown that the HLLI Riemann solver provides the theoretical
minimum dissipation that is consistent with a stable numerical
scheme. In section 3, we demonstrate by explicit examples that the
HLLI RMHD Riemann solver preserves the Alfven wave with high
accuracy and respects the contact discontinuity. In Section 4, we
incorporate the important aspects of multi-dimensionality with the
MuSIC RMHD Riemann solver. This is truly a multi-dimensional scheme
and we demonstrate that its ability to resolve the strongly
interacting region substantially reduces Alfvenic dissipation
compared to higher dimensional schemes that utilize 1-D Riemann
solvers in each direction. In our final discussion section, we
describe how a numerical scheme that utilizes the MuSIC Riemann
solver would be suitable for specialized simulations that would shed
light on the causal physics of the time evolution of event horizon
magnetospheres and help define the full panoply of physically
allowed and disallowed solutions.

\section{Dissipation In Riemann Solvers}
In this section, we illustrate the mathematical implications of the
IACS in conservative upwind schemes that utilize Riemann solvers.
(Please also note that schemes that do not use Riemann solvers
necessarily have to introduce even higher levels of dissipation.
This is because they cannot discriminate between wave families in
the way that some of the better Riemann solvers can.) For simplicity
and without loss of generality, consider a one-dimensional grid. The
conservation law that must be solved in each direction and at each
time step can be formally written as
\begin{equation}
\frac{\partial U}{\partial x^{0"}} + \frac{\partial F}{\partial
x^{1"}} =0\;,
\end{equation}
For a fine enough mesh and a well behaved coordinate system, the
covariant derivatives can be replaced with ordinary derivatives in
the conservation equation. In fact, RMHD Riemann solvers are used in
GRMHD (general relativistic MHD) simulations
\cite{kom04,gam03,eti15}. In an integral (weak solution) solution of
the Riemann problem, the higher order corrections due to connection
coefficients will be small (bilinear) corrections compared to the
integral of the derivative terms which are linear in the space-time
mesh size. This is the essence of the validity of ignoring the
source (connection coefficient) terms in the GRMHD Riemann solvers
and is a manifestation of the equivalence principle. We note that in
the GRMHD conservation law the connection coefficient terms (source)
terms occur. The error induced by these terms can be made
arbitrarily small on a fine enough mesh compared to the differential
terms. However, in practice the mesh might be coarse enough that the
connection terms represent source terms that modify the solution of
the conservation law in each time step \cite{del07}. This is not
discussed further in this section which is concerned only with the
Riemann solvers in GRMHD. The present paper does not focus on a
consideration of stiff source terms.

\begin{figure}
\includegraphics[width=120 mm, angle= 0]{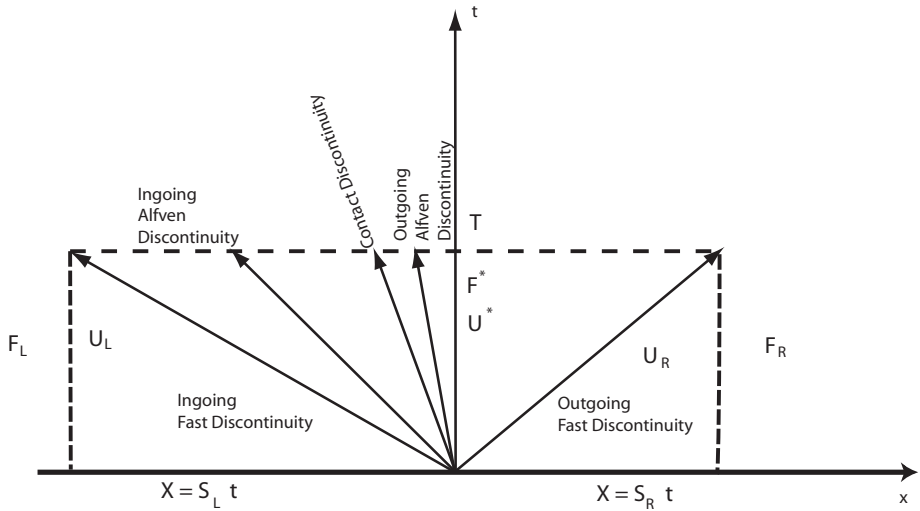}
\caption{An example of a one dimensional Reimann problem in which
the flow is super-Alfvenic inward (to the left). The flow is not
super-fast inward. This figure is used to illustrate what happens in
a Riemann problem at the interface between two cells in a numerical
scheme inside of the IACS. The 5-wave Riemann fan is illustrated
(the slow waves are ignored without loss of generality) in order to
show the difference of the resolved flux at the interface calculated
with a 2-wave HLL Riemann solver and the same calculation performed
with a 5-wave HLLI Riemann solver. Outgoing and ingoing are defined
in the frame of reference of the plasma.}
\end{figure}

\par Consider the nature of the flow at the IACS. We are especially interested in the
propagation of different RMHD wave families relative to the IACS
which, in principle, could be stationary relative to the
computational mesh. The space-time diagram is indicated in Figure 1.
The flow is super-Alfvenic inward (to the left). The flow is not
super-fast inward. This figure is a spacetime diagram of the MHD
characteristics at an interface between cells inside of the IACS.
The 5-wave Riemann fan is illustrated (the slow waves are ignored
without loss of generality) in order to show the difference in the
resolved flux that is produced by an HLL Riemann solver and a 5-wave
based Riemann solver. The spacetime in Figure 1 is split into
various zones. The world lines of isolated discontinuities that
emanate from the zone boundary are shown in Figure 1. The resolved
flux is the numerical flux in the zone that straddles the time axis.
Inside of the IACS, the resolved flux is the numerical flux in the
zone bounded on the left by Alfven wave that is outgoing in the
proper frame (but ingoing on the computational mesh) and on the
right by the outgoing fast wave. Physically, since the outgoing
Alfven wave overlies the time axis in Figure 1, we are interested in
capturing the stationary Alfvenic surface with maximum precision and
the least possible numerical diffusion.
\par Consider a one-dimensional Riemann problem with left and
right states $\mathbf{U}_{L}$ and $\mathbf{U}_{R}$ that are
separated by a Riemann fan with extremal speeds that span $[S_{L},\,
S_{R}]$. Let the fluxes that correspond to the left and right states
be given by $\mathbf{F}_{L}$ and $\mathbf{F}_{R}$. The numerical
flux from practically any one-dimensional Riemann solver can be
formally written as
\begin{equation}
\mathbf{F}^{*} = \frac{1}{2}(\mathbf{F}_{L} + \mathbf{F}_{R})
-\frac{1}{2}\mathbf{\Theta}(\mathbf{U}_{R} - \mathbf{U}_{L})\;.
\end{equation}
The first term, which is the average of the left and right fluxes in
the above expression, simply provides a centered, non-dissipative
flux. The second term in the above expression is known as the
dissipation term. The matrix, $\mathbf{\Theta}$, in the second term
is the viscosity matrix, it regulates the dissipation of the Riemann
solver. Further details on the ensuing mathematics can be found in
\cite{dum16}, Appendix B. Here we provide just enough results to
show the difference between the dissipation from the HLL Riemann
solver and the HLLI Riemann solver.
\par
The viscosity matrix is usually expressed in terms of the right and
left eigenvectors of the Roe matrix. Denoting the Roe matrix by
$\mathbf{A}(\mathbf{U}_{L}, \, \mathbf{U}_{R})$, we will refer to
its left and right eigenvectors by $\mathbf{L}$ and $\mathbf{R}$.
The eigenvalues of the Roe matrix will be denoted by a diagonal
matrix, $\mathbf{\Lambda}$. The viscosity matrix for the HLLI
Riemann solver can be written as
\begin{equation}
\mathbf{\Theta} =\mathbf{R}\mathbf{\Sigma}\mathbf{L}\,
\end{equation}
with
\begin{equation}
\mathbf{\Sigma} =\frac{S_{R} +S_{L}}{S_{R}-S_{L}}\mathbf{\Lambda}
-2\frac{S_{R}S_{L}}{S_{R}-S_{L}}\mathbf{I}
+2\frac{S_{R}S_{L}}{S_{R}-S_{L}}\mathbf{\delta}\;.
\end{equation}
Here $\mathbf{I}$ is the identity matrix and (for our purposes)
$\mathbf{\delta}$ is a special diagonal matrix that is introduced
into the HLLI Riemann solver to judiciously reduce dissipation.
Notice, therefore, that $\mathbf{\Sigma}$ is also a diagonal matrix.
Please observe from the previous equation that the viscosity matrix
introduces dissipation on a wave-by-wave basis, i.e. if the diagonal
term corresponding to a particular wave becomes zero, the
dissipation that is provided to that wave will also become zero. If
we choose the diagonal terms in $\mathbf{\delta}$ just right, we can
minimize the dissipation and even guarantee that the dissipation of
standing Alfven waves is exactly zero. This is exactly what has been
done in Dumbser and Balsara (2016). Those authors provide precise
expressions for the diagonal matrix, $\mathbf{\delta}$, which ensure
that stationary waves (whether they are Alfven waves or the entropy
wave) have zero dissipation. For the sake of completeness, we
catalogue their specification of the $\rm{i}^{\rm{th}}$ term of the
diagonal matrix, $\mathbf{\delta}$, as
\begin{equation}
\delta_{i} = 1-
\frac{\rm{min}(\lambda_{i},\,0)}{S_{L}}-\frac{\rm{max}(\lambda_{i},\,
0)}{S_{R}} \;.
\end{equation}
Here $\lambda_{i}$ is the $\rm{i}^{\rm{th}}$ eigenvalue of the Roe
matrix corresponding to the wave that we are interested in. We see
that the dissipation is finely tuned so that a moving Alfven wave
gets just the minimum amount of dissipation that it needs,
consistent with numerical stability. For example, a wave that
propagates slowly relative to the computational mesh is given
smaller dissipation compared to a wave that is propagating at high
speed on the mesh. This decision to regulate the dissipation
according to the wave speed is also what is demanded by numerical
stability.
\par The viscosity matrix for the HLL Riemann solver is retrieved
by setting $\mathbf{\delta} =0$. In that case, a standing Alfven
wave has non-zero dissipation which means that the Alfven waves at
the IACS surface will dissipate. Consequently, a numerical code that
is based on the HLL Riemann solver (especially if it is operated at
low to modest resolution) will not treat the IACS as a one-way
surface with respect to the propagation of Alfven waves. We feel
that this is a very important observation. Furthermore, with
$\mathbf{\delta} =0$, it is easy to see that the HLL Riemann solver
gives all waves a non-zero dissipation regardless of their wave
speed. To see this, let $\lambda_{i}$ be a specific eigenvalue. Then
the $\rm{i}^{\rm{th}}$ term for the diagonal matrix,
$\mathbf{\Sigma}$ can be written as .
\begin{equation}
\Sigma_{i} = \frac{S_{R}(\lambda_{i}-S_{L})-
S_{L}(S_{R}-\lambda_{i})}{S_{R}-S_{L}} \;.
\end{equation}
For the sub-sonic case shown in Figure 1 we have $S_{L} < 0 <
S_{R}$. We see that $\Sigma_{i} > 0$ for all intermediate
eigenvalues, $\lambda_{i}$, with $S_{L} < \lambda_{i} < S_{R}$.
Consequently all intermediate waves, like Alfven waves or contact
discontinuities, will always be dissipated by the HLL Riemann
solver.
\par In this section, we have only given a flavor of the
dissipation characteristics of the HLLI Riemann solver and how it
offers a substantial improvement over the HLL Riemann solver. The
reader who is interested in details should please read Dumbser and
Balsara (2016). The eigenvectors for RMHD that were used in this paper can all be obtained from \cite{bal01} or \cite{amm10} .

\section{One Dimensional Riemann Solvers in RMHD}

\par The Introduction has shown that it is very desirable to have Riemann solvers
that can capture stationary, isolated contact discontinuities as
well as stationary, isolated Alfven waves. Indeed the first of the
one-dimensional Riemann solvers for RMHD by \cite{kom99} and
\cite{bal01} were Roe-type Riemann solvers. Because such Riemann
solvers retain the entire set of eigenvectors for the RMHD system,
they can indeed capture stationary, isolated contact discontinuities
as well as stationary, isolated Alfven waves on a mesh. There has
been a recent effort by \cite{amm10} to revive the use of Roe-type
Riemann solvers in RMHD simulations, but the effort has met with
limited success owing to the exorbitant computational cost of such
Riemann solvers, especially for RMHD. The Roe-type Riemann solvers
also have an inherent deficiency. This has to do with their loss of
positivity of density and pressure in certain circumstances
\cite{ein88,ein91}. It is easy to find mentions in the early
literature on RMHD, \cite{kom99,kom04,kom05}, showing that the early
RMHD simulation codes struggled with positivity issues and,
therefore, reverted to the use of the HLL Riemann solver. This issue
is relevant to the extremely low density environment of event
horizon magnetospheres.

\par HLLD Riemann solvers. \cite{mig09}, also enable a code
to capture stationary, isolated contact discontinuities as well as
stationary, isolated Alfven waves. But they have their own set of
attendant problems, as discussed in the Introduction. HLLC Riemann
solvers, \cite{mb06,hj07,kb14}, represent a compromise position
where they enable a code to capture stationary, isolated contact
discontinuities, but not Alfven waves. The HLLI Riemann solver of
\cite{dum16} is built on top of an HLL Riemann solver, so it
inherits all the beneficial positivity properties of the HLL Riemann
solver. However, it introduces sub-structure in the Riemann fan.
Typically, that substructure includes the contribution from
eigenvectors of the contact discontinuity and eigenvectors
associated with Alfven waves. The eigenvectors for the fast and slow
magnetosonic waves are very expensive to evaluate computationally,
and their evaluation is avoided in the HLLI Riemann solver. As a
result, the HLLI Riemann solver enables a code to capture
stationary, isolated contact discontinuities as well as stationary,
isolated Alfven waves at a very low computational cost. Unlike HLLC
and HLLD, the HLLI Riemann solver does not require an iterative
solution, thereby ensuring that it has even lower computational
cost. In the next few paragraphs we demonstrate this facet of the
HLLI Riemann solver for RMHD.
\begin{figure}
\centering{
\includegraphics[width=80 mm, angle= 0]{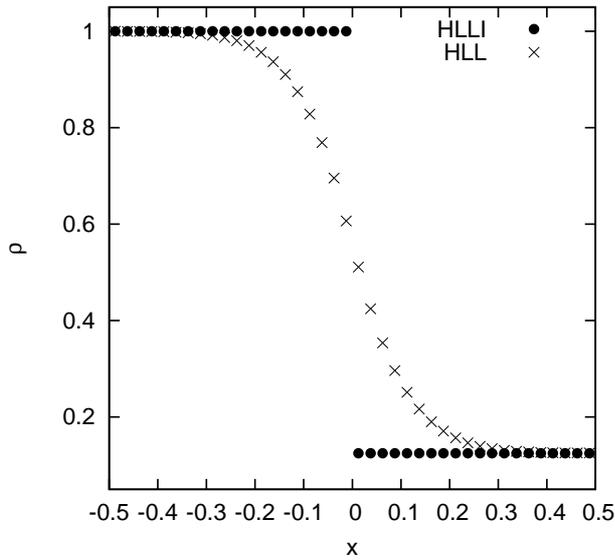}}
\caption{Figure 2 shows two simulations of an isolated, stationary
contact discontinuity. The density variable is shown. The result
from the HLL Riemann solver is shown with crosses while the result
from the HLLI Riemann solver is shown with dots.}
\end{figure}

\par Figure 2 shows two simulations of an isolated, stationary contact discontinuity as suggested
by \cite{hj07}. The density variable is shown. We use the same
parameters as the previous authors and we run the simulation to a
final time that is ten times larger than the one suggested by
Honkkila and Janhunen. The result from the HLLI Riemann solver is
shown with filled dots, the result from the HLL Riemann solver is
shown with crosses. We see that the HLL Riemann solver has produced
significant dissipation of the contact discontinuity, while the HLLI
Riemann solver has captured the contact discontinuity exactly.

\begin{figure}
\centering{
\includegraphics[width=80 mm, angle= 0]{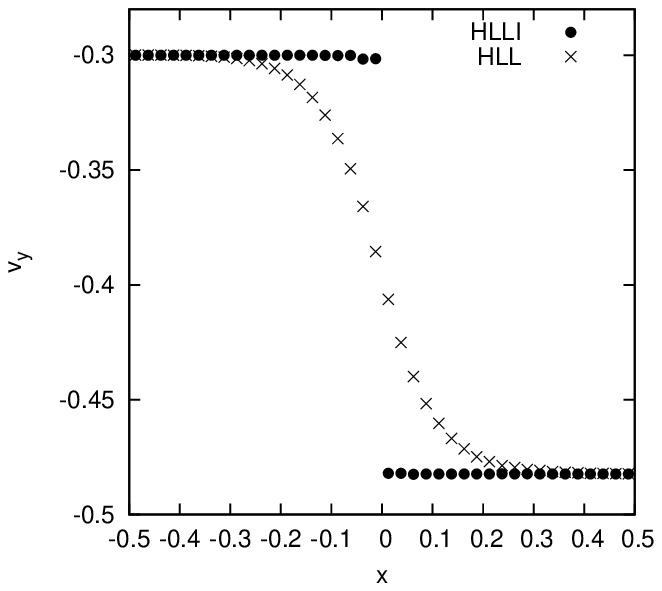}
\includegraphics[width=80 mm, angle= 0]{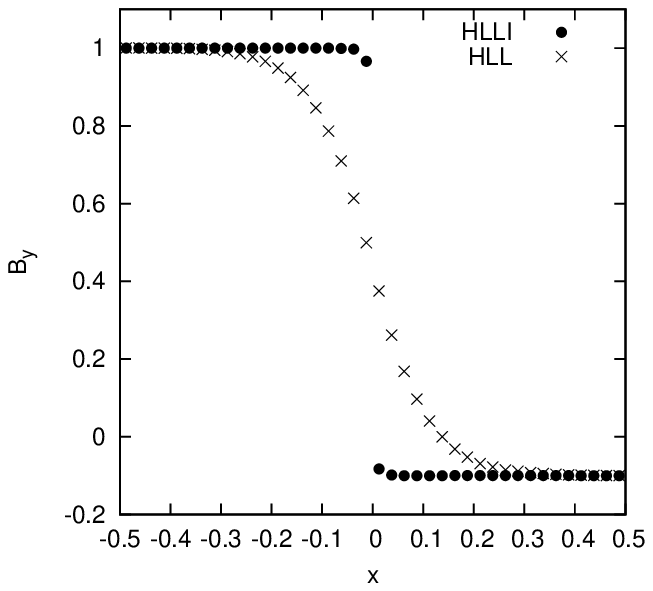}}
\caption{Figure 3 shows two simulations of an isolated, stationary
Alfven wave. The transverse velocity and magnetic field are shown.
The result from the HLL Riemann solver is shown with crosses while
the result from the HLLI Riemann solver is shown with dots.}
\end{figure}

\par Figure 3 shows two simulations of an isolated, stationary Alfven wave. The transverse
velocity and magnetic field are shown. This problem was suggested by
\cite{mig09} and we use the same parameters as the previous authors
but we run the simulation to a final time that is four times larger
than the final time quoted by Mignone, Ugliano and Bodo (2009). The
result from the HLLI Riemann solver is shown with filled dots, the
result from the HLL Riemann solver is shown with crosses. As before,
we see that the HLL Riemann solver has produced significant
dissipation of the Alfven wave discontinuity, while the HLLI Riemann
solver has captured the Alfven wave discontinuity exactly. Figures 2
and 3 both used a standard, second order scheme, the only difference
being the use of the HLLI Riemann solver.

\section{Multidimensional Propagation of Alfven Waves}

\par Figures 2 and 3 showed that the one-dimensional HLLI Riemann solver can dramatically
reduce the dissipation compared to the HLL Riemann solver.
Multidimensional treatment of Alfven waves on a computational mesh
requires a multidimensional Riemann solver. In \cite{bal04}, we were
able to formulate a test problem that measures the ability of a
multidimensional MHD code to propagate Alfven waves with the least
amount of dissipation. It has been shown that this test problem is
very important in benchmarking the dissipation characteristics of
multidimensional codes for classical MHD. The analogous problem
which benchmarks the low dissipation propagation of Alfven waves in
RMHD has been recently presented in Balsara and Kim (2016). It
consists of torsional Alfven waves propagating obliquely to the mesh
lines of a two-dimensional mesh. The mesh has 120x120 zones. We do not repeat the problem
description. Instead, we show the results and intercompare with
older methods that involve dissipation doubling from Gardiner and
Stone (2005, 2008).

\begin{figure}
\centering{
\includegraphics[width=80 mm, angle= 0]{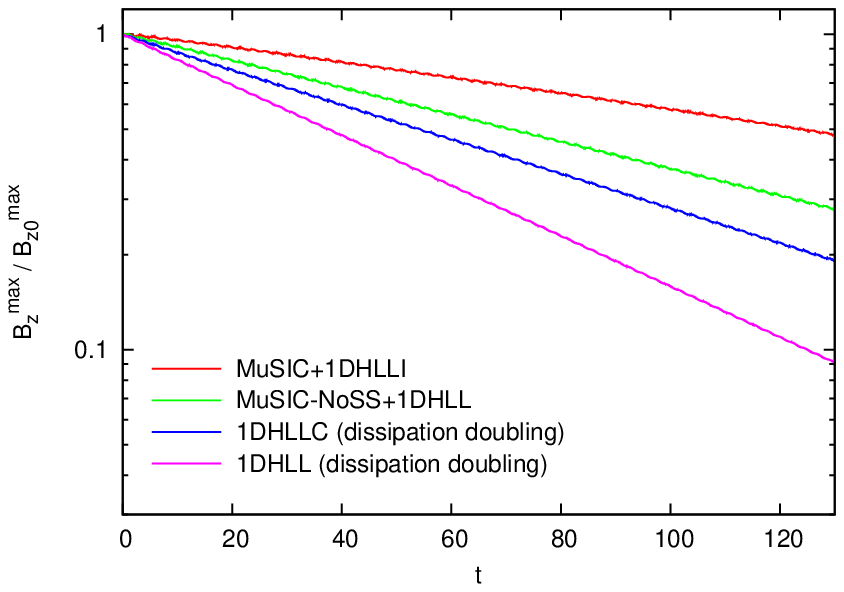}
\includegraphics[width=80 mm, angle= 0]{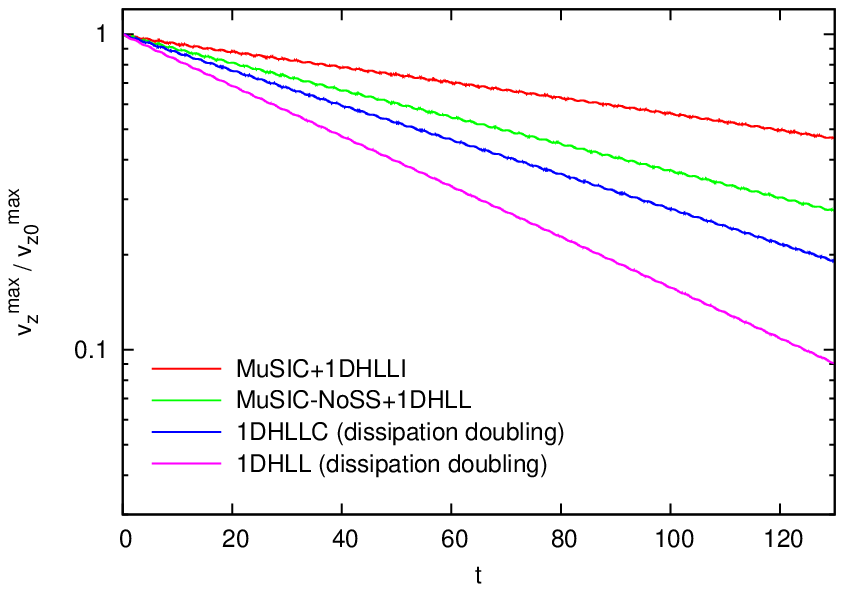}}
\caption{Figure 4 shows the results of the torsional Alfven wave
dissipation test. A second order WENO reconstruction was used in all
these tests. Figure 4a shows the decay in the z-component of the
velocity as a function of time. Figure 4b shows the same for the
z-component of the magnetic field.}
\end{figure}

\par Figure 4 shows the results of the torsional Alfven wave dissipation test from Balsara and Kim (2016).
In Figure 4, we use the same second order reconstruction algorithm
from the RMHD version of the RIEMANN code. Figure 4a shows the decay
in the z-component of the velocity of the Alfven wave as a function
of time. Figure 4b shows the same for the z-component of the
magnetic field of the Alfven wave. The vertical axis is
logarithmically scaled. A faster rate of decline in Figure 4
indicates that the associated numerical scheme has higher
dissipation. The curve that is labeled ``MuSIC+1DHLLI'' uses the
one-dimensional HLLI Riemann solver at the zone faces and the MuSIC
Riemann solver with sub-structure at the zone edges. We see that it
displays minimal dissipation. This is because the MuSIC Riemann
solver is designed to be the exact, multidimensional analogue of the
one-dimensional HLLI Riemann solver. The curve that is labeled
``MuSIC-NoSS+1DHLLI'' uses the same algorithmic combination with one
simple exception. The MuSIC Riemann solver is prevented from
endowing sub-structure to the strongly-interacting state. We see
that when the sub-structure in the MuSIC Riemann solver is
artificially removed, the dissipation of Alfven waves increases.
This makes the very nice point that all facets of the newly designed
MuSIC Riemann solver play a role in reducing dissipation. It is very
useful to cross-compare with the dissipation doubling ideas from
\cite{gs05,gs08}. The curve that is labeled ``1D HLLC (dissipation
doubling)'' doubles the dissipation in the HLLC Riemann solver using
the ideas from \cite{gs05,gs08}. Despite the one-dimensional HLLC
Riemann solver being an able performer, we see that it dramatically
increases the dissipation that is provided to the torsional Alfven
waves. Lastly, one is most interested in understanding what happens
when the dissipation doubling ideas from \cite{gs05,gs08} are
applied to the one-dimensional HLL Riemann solver. This is shown in
Figure 4 by the curve labeled ``1D HLL (dissipation doubling)''. We
see that Alfven waves are strongly dissipated.

\begin{figure}
\centering{
\includegraphics[width=80 mm, angle= 0]{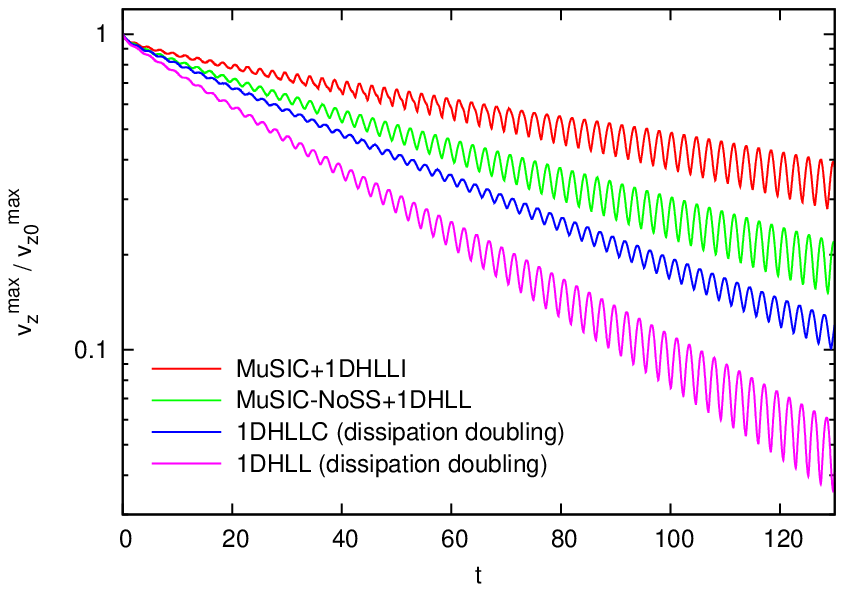}
\includegraphics[width=80 mm, angle= 0]{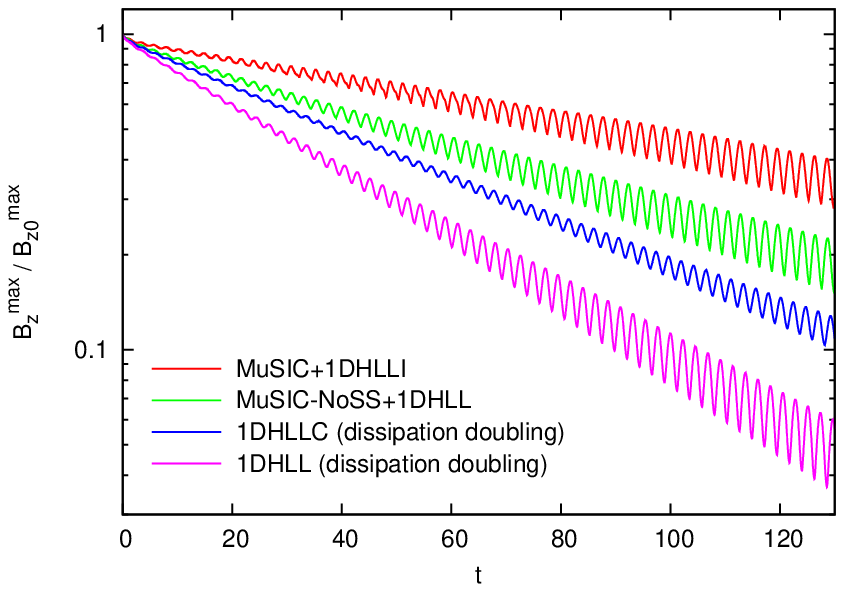}}
\caption{Figure 5 is analogous to Figure 4 with the exception than
an MC limiter was used. The MC limiter is considered inferior to a
good WENO scheme. Comparing Figures 4 and 5, this observation is
apparent in the figures.}
\end{figure}

\par It should also be emphasized that the reconstruction that
was used in Figure 4 is the linear part of the r=3 WENO
reconstruction. This is already a very superior reconstruction
strategy. It is almost as superior as a true third order
reconstruction strategy. It is quite possible that reconstruction is
done with a second order TVD limiter, like the MC limiter. In that
case, the analogous results are shown in Figure 5. We see
considerably increased dissipation in Figure 5 compared to Figure 4.

\begin{figure}[ht]
\centerline{
\includegraphics[width=80 mm, angle= 0]{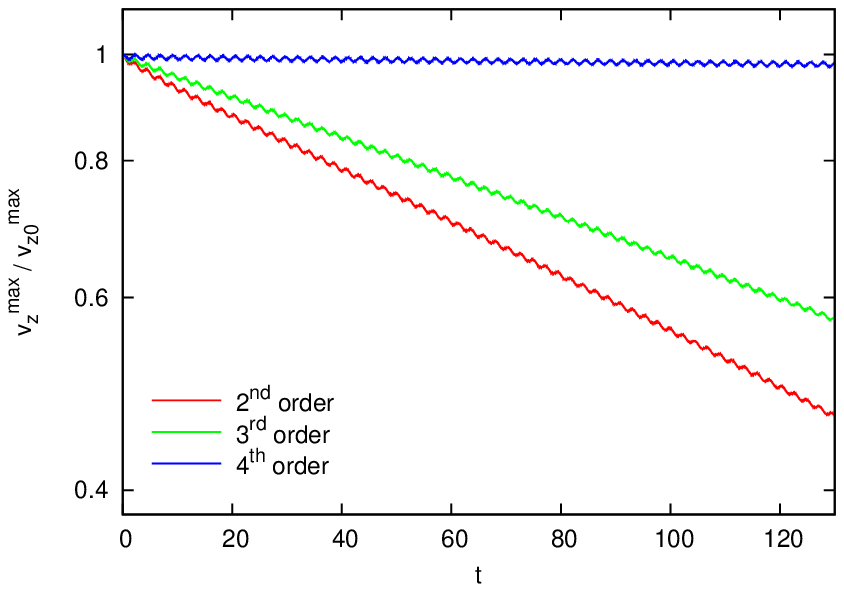}
\includegraphics[width=80 mm, angle= 0]{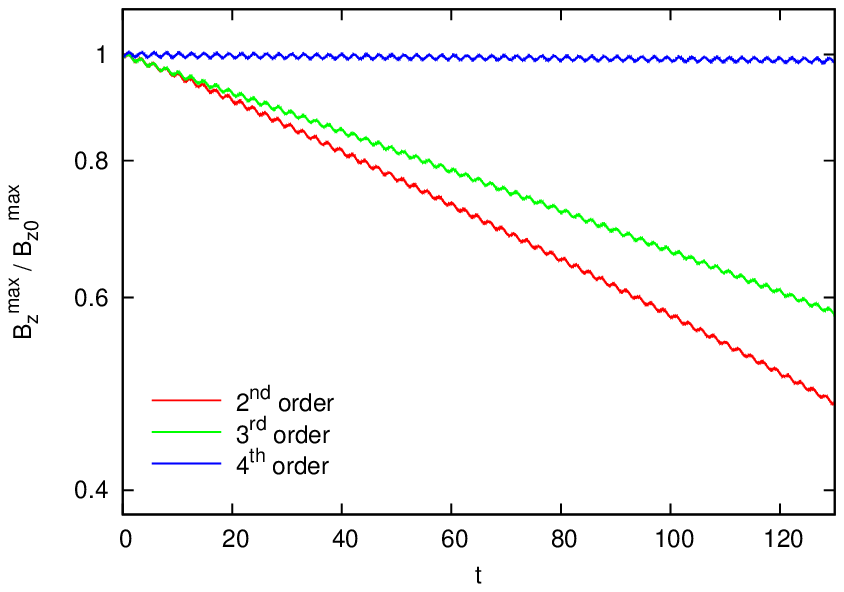}}
\caption{Figure 6 shows the same Alfven wave propagation test. This
time, we used different ADER-WENO schemes with increasing order of
accuracy. We also used the 1D HLLI Riemann solver along with the 2D MuSIC Riemann solver with sub-structure. We see that higher order schemes produce lower
dissipation.}
\end{figure}

\par In \cite{bk16}, ADER-WENO schemes were designed that go all the way up to fourth order
of accuracy. It is, therefore, very interesting to ask whether
improved accuracy gives us an improved result for the propagation of
torsional Alfven waves. Figure 6 shows the propagation of torsional
Alfven waves when second, third and fourth order ADER-WENO schemes
are used. All these schemes used the one-dimensional HLLI Riemann
solver at the zone faces and the MuSIC Riemann solver with
sub-structure at the zone edges. We clearly see that the higher
order schemes show vastly reduced dissipation. In \cite{bk16}, we
also demonstrate that modern high order schemes perform robustly
even in the vicinity of strong shocks. Thus the barrier to their use
in astrophysics is dramatically reduced by this work.

\begin{figure}[ht]
\centerline{
\includegraphics[width=80 mm, angle= 0]{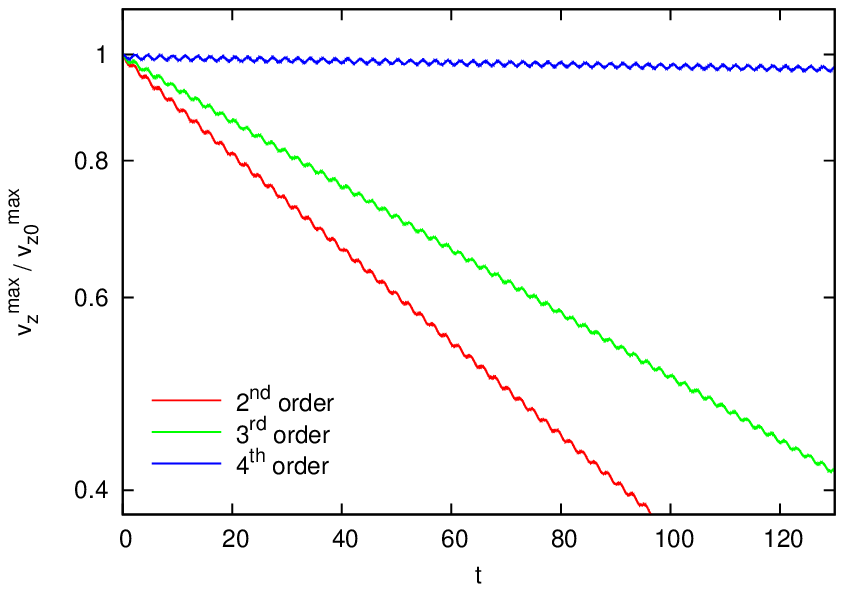}
\includegraphics[width=80 mm, angle= 0]{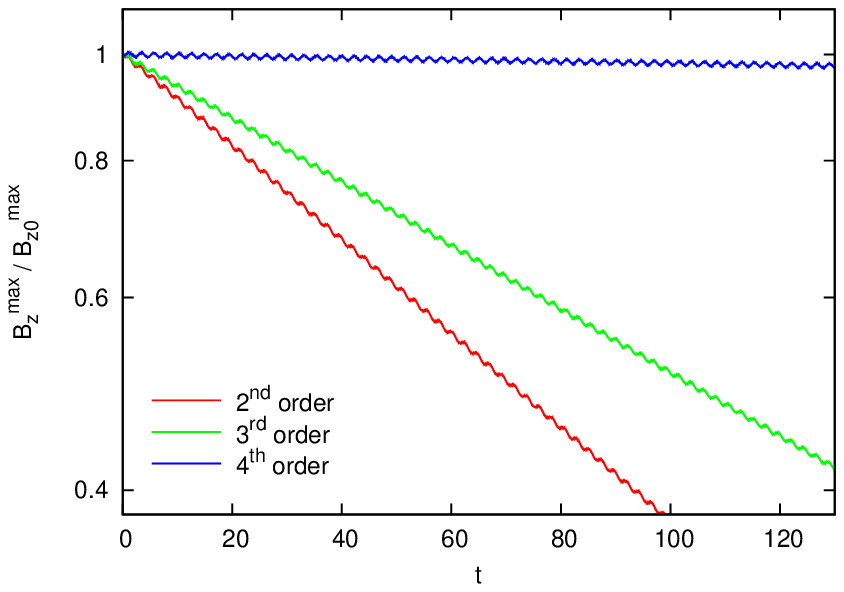}}
\caption{Figure 7 shows the same Alfven wave propagation test as Figure 6 but when a lower quality Riemann solver is used.
We used the same ADER-WENO schemes as in Figure 6. The only difference is that the simulations in this
Figure were run with a 1D HLL Riemann solver and a 2D HLL Riemann solver. Comparing Figure 6 to Figure 7 enables
us to appreciate the improvement that Riemann solvers with sub-structure provide in reducing dissipation.}
\end{figure}

\par Figure 6 clearly shows us that the combination of a high accuracy method and 1D and 2D Riemann solvers that preserve sub-structure produces very low dissipation. This is especially apparent in the fourth order simulation shown in Figure 6. Out of curiosity, we can always ask what fraction of the improvement in Alfven wave propagation stems from the use of a higher order scheme and what fraction of the improvement in Alfven wave propagation stems from the use of Riemann solvers that retain substructure? For that reason, the same simulations from Figure 6 were run again with 1D and 2D Riemann solvers that do not preserve sub-structure. The results are shown in Figure 7. In other words, for Figure 7, the 1D Riemann solver was an HLL Riemann solver and the 2D Riemann solver was a 2D analogue of an HLL Riemann solver. Consequently, Figure 7 shows the result of Alfven wave propagation when lower quality Riemann solvers are used. We see that the second order result in Figure 7 is substantially more dissipative than the second order result from Figure 6. We also see that the second order result from Figure 6 has a dissipation that is comparable to the third order result from Figure 7. In other words, using a Riemann solver with sub-structure produces a very palpable improvement in second and third order schemes. When we compare the fourth order results from Figures 6 and 7, we see that they are indeed quite comparable. In other words, we suggest that at fourth and higher orders of accuracy the value of a Riemann solver that preserves sub-structure is diminished because the fourth order reconstruction itself is so very accurate! Note though that a third order scheme will typically be two to three times more expensive compared to a second order scheme. Similarly, a fourth order accurate scheme will be about three times more expensive compared to a third order scheme. For that reason, it is very profitable to try and extract as much performance and quality from a lower order scheme, especially if one does not have access to a higher order scheme.

\par Figures 6 and 7 show that there is always a small wiggle in the maximum amplitude of the Alfven waves. This wiggle stems from the fact that the Alfven waves in our test problem have large amplitude and are, therefore, prone to small initialization errors or errors in start-up transients that never go away. The non-relativistic analogue of this test problem also shows the same issue. The wiggles make the test problem used in Figure 6 an inadequate test problem for demonstrating order of accuracy. That is especially true at fourth order where the wiggles have an amplitude that is comparable to the decay of the Alfven wave. (In \cite{bk16} we do present better test problems for demonstrating the accuracy of an RMHD scheme.) However, please observe from Fig. 6 that at second and third orders of accuracy, the rate of decay in the Alfven wave amplitude is much larger than the wiggles, so that it can be used to document the second and third orders of accuracy of the schemes used here. Table 1 shows the $L_{1}$ error in the z-momentum density and the z-component of the magnetic field as a function of mesh resolution at the second order. Table 2 shows the $L_{1}$ error in the z-momentum density and the z-component of the magnetic field as a function of mesh resolution at the third order. The errors are shown at a time of 30 units, by which point the decay in the Alfven waves at second and third orders of accuracy is much larger than the amplitude of the wiggles. We see that the second and third order accurate methods do indeed achieve their design accuracies.

\begin{table}
\begin{tabular}{ c c c c c}
\hline
zones & $L_1$ error & Accuracy & $L_1$ error & Accuracy \\
& z-momentum density & & z-magnetic field & \\
\hline
$60\times60$ & $2.7693 \times 10^{-1}$ & & $1.1971 \times 10^{-1}$ & \\
$120\times120$ & $7.5660 \times 10^{-2}$ & 1.87 & $3.3246 \times 10^{-2}$ & 1.85\\
$240\times240$ & $1.6083 \times 10^{-2}$ & 2.23 & $7.1989 \times 10^{-3}$ & 2.21\\
\hline
\end{tabular}
\caption {$L_{1}$ error in the z-momentum density and the z-component of the magnetic field as a function of mesh resolution at the second order}
\end{table}

\begin{table}
\begin{tabular}{ c c c c c}
\hline
zones & $L_1$ error & Accuracy & $L_1$ error & Accuracy \\
& z-momentum density & & z-magnetic field & \\
\hline
$60\times60$   & $2.1295 \times 10^{-1}$ &      & $9.2981 \times 10^{-2}$ & \\
$120\times120$ & $3.7593 \times 10^{-2}$ & 2.50 & $1.6720 \times 10^{-2}$ & 2.48\\
$240\times240$ & $5.1524 \times 10^{-3}$ & 2.87 & $2.2025 \times 10^{-3}$ & 2.92\\
\hline
\end{tabular}
\caption {$L_{1}$ error in the z-momentum density and the z-component of the magnetic field as a function of mesh resolution at the third order}
\end{table}

\section{Discussion and Future Prospects}
In this article, we describe subtle points uniquely associated with
numerical simulations of event horizon magnetospheres. All inflows
must pass through an IACS, thereby rendering the interior region out
of Alfven wave communication with the out-flowing wind. Thus, the
IACS is a charge horizon (a one-way membrane) for the global flow
since the Alfven wave is the only wave that carries a physical
charge \cite{pun05}. The IACS causally excises most of the active
region of space-time (the ergosphere) that would be the putative
element to enforce rapid rotation of the system. The implication is
that simulations of event horizon magnetospheres should be performed
with numerical schemes that represent the Alfvenic properties of the
system with high fidelity.
\par It was demonstrated  that an improvement of numerical accuracy can
be attained by utilizing numerical schemes based on HLLI and MuSIC
Riemann solvers rather than schemes based on 1-D HLL solvers. In
particular, for 1-D RMHD solvers, the HLLI Riemann solver provides
the theoretical minimum dissipation of the Alfven wave and contact
discontinuities that can still ensure a numerically stable scheme.
In Section 4, we discussed the multidimensional extension of HLLI
RMHD Riemann solver; the RMHD MuSIC Riemann solver. It is in higher
dimensions that we see an even larger improvement over schemes based
on 1-D HLL solvers. The MuSIC Riemann solver was shown to
significantly reduce the dissipation of Alfven waves in large part
to its ability to resolve the strongly interacting region that is
typically ignored in schemes based on 1-D HLL Riemann solvers. We also show that very high order schemes might be free of the excessive dissipation that arises from a lower quality Riemann solver. However, that transition occurs only when schemes of fourth and higher order of accuracy are used.

\par Low Alfven and contact discontinuity dissipation in a numerical scheme, such as those based on MuSIC,
should allow the proper propagation of Alfven wave information in
the following unique circumstances endemic to event horizon
magnetospheres that were discussed in the Introduction.

\begin{enumerate}
\item At the risk of being repetitive, the paired wind systems evolves outward and inward towards two
asymptotic infinities as opposed to having a causal MHD boundary at
one terminus. Thus, unlike other MHD wind problems there is a more
complex critical point structure. Most specifically, all inflows
pass through the inner Alfven critical surface. Thereby causally
disconnecting the outflowing wind from Alfven radiation emanating
from the majority of the rapidly rotating ergospheric plasma. The
Alfven wave is the only wave that propagates a physical charge and
thus should be involved in the establishment of the Goldreich-Julian
charge density or equivalently, the field line rotation rate,
$\Omega_{F}$. Thus, reducing the numerical dissipation of the Alfven
wave by implementing the HLLI or MuSIC Riemann solvers would seem to
to help in this regard.
\item The paired wind system constantly drains itself of plasma in
the MHD limit. Thus, plasma injection by means of a mass floor is
required. This process will dissipate MHD waves generated by the MHD
system and inject new MHD waves. The process is nontrivial and has
been shown to modify $\Omega_{F}$ significantly in certain 3-D
simulations. Since, in principle, it can modify $\Omega_{F}$ and
mass injection perturbs the Alfven waves generated in the system, a
Riemann solver lowers Alfven dissipation might shed light on the
nature of the transients that occur as different plasma injection
scenarios are explored.
\item Another related issue is the large numerical diffusion of
plasma from the bounding accretion disk into the event horizon
magnetosphere. Diffusion is substantially reduced if the Riemann
solver respects the contact discontinuity. Since the source of
plasma injection might be important to the establishment of
$\Omega_{F}$ this is an important issue as well. Riemann solvers,
such as the HLLI or MUSIC Riemann solvers, which treat contact
discontinuities explicitly can help in this regard.

\end{enumerate}
\par In this article, we presented both theoretical and numerical
arguments that support the notion that numerical schemes based on
the multi-dimensional MuSIC Riemann solver can potentially provide
an improvement over existing methods of modeling 3-D event horizon
magnetospheres. In particular, the MuSIC Riemann solver is both well
suited for the low density environment endemic to the event horizon
magnetosphere and it provides reduced dissipation of the Alfven and
contact discontinuities. It is also computationally efficient which
offsets the cost of improving the computational accuracy.

\par In \cite{bk16} we provide a subluminal scheme for RMHD
as well as a discussion of the MuSIC Riemann solver. The reader who
wishes to get further information can also visit the second author's
website http://www.nd.edu/~dbalsara/Numerical-PDE-Course.

\begin{acknowledgements}
DSB acknowledges support via NSF grants NSF-ACI-1307369,
NSF-DMS-1361197 and NSF-ACI-1533850. DSB also acknowledges support
via NASA grant NASA-NNX 12A088G. Several simulations were performed
on a cluster at UND that is run by the Center for Research
Computing. BP thanks ICRANet for support.
\end{acknowledgements}
\section{Competing Interests} The authors declare that they have no competing interests
\section{Contributions Section} BP conceived of the study and
provided the contextual information on black holes. DB, J.K. and
S.G. wrote most of the manuscript and provided the numerical
simulations.


\begin{thebibliography}{999}
\bibitem[Ant{\'o}n et al.(2010)]{amm10} Ant{\'o}n, L., Miralles, J.~A., Mart{\'{\i}}, J.~M., et al.
Relativistic Magnetohydrodynamics: Renormalized Eigenvectors and
Full Wave Decomposition Riemann Solver ApjS 188 1-31 (2010)
\bibitem[Balsara (2001)]{bal01} Balsara, D. Total Variation Diminishing Scheme for Relativistic Magnetohydrodynamics ApJS, 132,
83-101 (2001)
\bibitem[Balsara(2004)]{bal04} Balsara, D.~S. Second-Order-accurate Schemes for Magnetohydrodynamics with Divergence-free Reconstruction ApJS, 151, 149-184 (2004)
\bibitem[Balsara(2010)]{bal10} Balsara, D.~S. Multidimensional Extension of  the HLLE Riemann Solver; Application to Euler and Magnetohydrodynamical Flows, Journal of Computational Physics, 229,
1970-1983 (2010)
\bibitem[Balsara(2012)]{bal12} Balsara, D.~S. A two-dimensional HLLC Riemann solver for conservation laws: Application to Euler and magnetohydrodynamic flows Journal of Computational Physics, 231,
7476-7503 (2012)
\bibitem[Balsara(2014)]{bal14} Balsara, D.~S. Multidimensional Riemann Problem with Self-Similar Internal Structure – Part I –  Application to Hyperbolic Conservation Laws on Structured Meshes, Journal of Computational Physics, 277,
163-200 (2014)
\bibitem[Balsara et al.(2014)]{bda14} Balsara, D.~S., Dumbser, M., \& Abgrall, R. Multidimensional HLL and HLLC Riemann
S olvers for Unstructured Meshes – With Application to Euler and MHD
Flows Journal of Computational Physics, 261, 172-208 (2014)
\bibitem[Balsara(2015)]{bal15} Balsara, D.~S. Three dimensional HLL Riemann solver for conservation laws on structured meshes; Application to Euler and magnetohydrodynamic flows  Journal of Computational Physics, 295,
1-23 (2015)
\bibitem[Balsara \& Dumbser(2015)]{bd15} Balsara, D.~S., \& Dumbser, M. Multidimensional Riemann solvers, Conservation laws, Higher order Godunov schemes, Unstructured meshes, Euler flow, MHD Journal of Computational Physics, 287,
269-282 (2015)
\bibitem[Balsara \& Kim(2016)]{bk16} Balsara, D.~S., \& Kim, J. A subluminal relativistic magnetohydrodynamics scheme with ADER-WENO predictor and multidimensional Riemann solver-based corrector Journal of Computational
Journal of Computational Physics, 312, 357-384 (2016)
\bibitem[Balsara et al.(2016a)]{bvg16} Balsara, D.~S., Vides, J., Gurski, K., et al. A two-dimensional Riemann solver with self-similar sub-structure - Alternative formulation based on least squares projection Journal of Computational Physics, 304,
138-161 (2016)
\bibitem[Balsara et al.(2016b)]{bnd16} Balsara, D.~S., Nkonga, B., Dumbser, M., \& Munz, C.~D.\ 2016b, in preparation
\bibitem[Beckwith \& Stone(2011)]{bs11} Beckwith, K., \& Stone, J.~M. A Second-order Godunov Method for Multi-dimensional Relativistic Magnetohydrodynamics ApJS 193,
6-35 (2011)
\bibitem[Beskin and Zheltoukhov,(2013)]{bes13}  Beskin, V. S. and Zheltoukhov, A. A. On the structure of the magnetic field near a black hole in active galactic nuclei Astron. Lett., 39,
215-220 (2103)
\bibitem[De Villiers et al.(2003)]{dev03} De Villiers, J-P., Hawley, J., Krolik, Magnetically Driven Accretion Flows in the Kerr Metric. I. Models and Overall Structure ApJ 599
1238-1253 (2003)
\bibitem[Del Zanna et al.(2007)]{del07} Del Zanna, L., Zanotti, O., Bucciantini, N., \& Londrillo, P.ECHO: a Eulerian conservative high-order scheme for general relativistic magnetohydrodynamics and magnetodynamics, A \$ A, 473,
11-30 (2007)
\bibitem[Dumbser and Balsara (2016)]{dum16} Dumbser, M., Balsara, D., A New, Efficient Formula tion of the HLLEM Riemann Solver for General
Conservative and Non - Conservative Hyperbolic Systems J, Comp.Phys.
304 275 -319 (2016)
\bibitem[Einfeldt(1988)]{ein88} Einfeldt, B. On Godunov-Type Methods for Gas Dynamics SIAM J. Numer. Anal., 25
294-318 (1988)
\bibitem[Einfeldt et al.(1991)]{ein91} Einfeldt,, B. Munz, C-D. Roe, P \& Sjogreen, B. On Godunov-type
methods near low densities. J. Comput. Phys., 92 273-295 (1991)
\bibitem[Etienne et al.(2015)]{eti15} Etienne, Z., Paschalidis, V., Haas, R., Mösta, P., Shapiro, S., IllinoisGRMHD: an open-source, user-friendly GRMHD code for dynamical spacetimes Clas. and Quan. Gra.
32 5009 (2015)
\bibitem[Fragile et al(2007)]{fra07}Fragile, P. C., Blaes, O. M., Anninos, P., Salmonson, J. D. Global General Relativistic Magnetohydrodynamic Simulation of a Tilted Black Hole Accretion Disk ApJ 668
417-429 (2007)
\bibitem[Gammie et al(2003)]{gam03} Gammie C.F., McKinney J.C., Toth G, HARM: A Numerical Scheme for General Relativistic Magnetohydrodynamics ApJ, 589,
444-457 (2003)
\bibitem[Gardiner \& Stone(2005)]{gs05} Gardiner, T.~A., \& Stone, J.~M.An unsplit Godunov method for ideal MHD via constrained transport Journal of Computational Physics, 205,
509-539 (2005)
\bibitem[Gardiner \& Stone(2008)]{gs08} Gardiner, T.~A., \& Stone, J.~M.An unsplit Godunov method for ideal MHD via constrained transport in three dimensions Journal of Computational Physics, 227,
4123-4141 (2008)
\bibitem[Giacomazzo \& Rezzolla(2006)]{gia06} Giacomazzo, B., \& Rezzolla, L. The exact solution of the Riemann problem in relativistic magnetohydrodynamics Journal of Fluid Mechanics, 562,
223-259 (2006)
\bibitem[Giacomazzo and Rezzolla(2007)]{gia07} Giacomazzo B., Rezzolla L., WhiskyMHD: a new numerical code for general relativistic magnetohydrodynamics Class.Quant.Grav., 24(12),
S235-S258 (2007)
\bibitem[Hawley and Krolik(2006)]{haw06} Hawley, J., Krolik, K. Magnetically Driven Jets in the Kerr Metric ApJ 641
103-116 (2006)
\bibitem[Honkkila \& Janhunen(2007)]{hj07} Honkkila, V., \& Janhunen, P.HLLC solver for ideal relativistic MHD Journal of Computational Physics, 223,
643-656 (2007)
\bibitem[Kappeli \& Mishra(2014)]{km14} Kappeli, R. and Mishra, S.  Journal of Computational Physics, Well-balanced schemes for the Euler equations with gravitation 259,
199-219 (2014)
\bibitem[Kappeli \& Mishra(2016)]{km16} Kappeli, R. and Mishra, S. A well-balanced finite volume scheme for the Euler equations with gravitation. The exact preservation of hydrostatic equilibrium with arbitrary entropy stratification A \& A,
587 A94 -A110 (2016)
\bibitem[Kim \& Balsara(2014)]{kb14} Kim, J., \& Balsara, D.~S.A Stable HLLC Riemann solver for Relativistic
Magnetohydrodynamics Journal of Computational Physics, 270, 634-639
(2014)
\bibitem[Komissarov(1999)]{kom99}Komissarov, S. A Godunov-type scheme for relativistic magnetohydrodynamics MNRAS. 303
343-366 (1999)
\bibitem[Komissarov(2004b)]{kom04}Komissarov, S. General relativistic magnetohydrodynamic simulations of monopole magnetospheres of black holes MNRAS 350,
1431-1436 (2004)
\bibitem[Komissarov(2005)]{kom05}Komissarov, S. Observations of the Blandford-Znajek process and the magnetohydrodynamic Penrose process in computer simulations of black hole magnetospheres MNRAS 359,
801-808 (2005)
\bibitem[Krolik et al(2005)]{kro05} Krolik, K., Hawley, J., Hirose, S.
Magnetically Driven Accretion Flows in the Kerr Metric. IV.
Dynamical Properties of the Inner Disk ApJ 622, 1008 -1023 (2005)
\bibitem[McKinney and Blandford(2009)]{mck09}McKinney, J., Blandford, R. Stability of relativistic jets from rotating, accreting black holes via fully three-dimensional magnetohydrodynamic simulations MNRAS Letters 394
126-130 (2009)
\bibitem[McKinney et al.(2012)]{mck12}McKinney, J., Tchekhovskoy, A., Blandford, R. General relativistic magnetohydrodynamic simulations of magnetically choked accretion flows around black holes MNRAS 423
3083-3117 (2012)
\bibitem[Mignone \& Bodo(2006)]{mb06} Mignone, A., \& Bodo, G.An HLLC Riemann solver for relativistic flows - II. Magnetohydrodynamics MNRAS, 368,
1040-1054 (2006)
\bibitem[(Mignone, Ugliano and Bodo(2009)]{mig09}Mignone, A., Ugliano, M.; Bodo, G. A five-wave Harten-Lax-van Leer Riemann solver for relativistic magnetohydrodynamics MNRAS 393,
1141-1156 (2009)
\bibitem[Par{\'e}s(2006)]{par06} Par{\'e}s, C., Numerical methods for nonconservative hyperbolic systems: a theoretical framework SIAM J. Numer. Anal., 44,
300-321 (2006)
\bibitem[Punsly(2004)]{pun05} Punsly, B.Fast-Wave Polarization, Charge Horizons, and the Time Evolution of Force-free Magnetospheres, ApJL 612,
41-44 (2004)
\bibitem[Punsly(2008)]{pun08} Punsly, B. Black Hole Gravitohydromagnetics, second edition Springer-Verlag, New
York (2008)
\bibitem[White \& Stone(2015)]{ws15} White, C.~J., \& Stone, J.~M.GRMHD in Athena++ Using Advanced Riemann Solvers and Staggered-Mesh Constrained Transport
 (2015) arXiv:1511.00943
\bibitem[Zanotti \& Dumbser(2016)]{zd16} Zanotti, O., \& Dumbser, M.Efficient conservative ADER schemes based on WENO reconstruction and space-time predictor in primitive variables Computational Astrophysics and Cosmology, 3,
\#1 (2016) DOI: 10.1186/s40668-015-0014-x
\end{thebibliography}


\section{Figure Legends}
\begin{enumerate}
\item Super Alfvenic Riemann Fan. An example of a one dimensional Reimann problem in which
the flow is super-Alfvenic inward (to the left). The flow is not
super-fast inward. This figure is used to illustrate what happens in
a Riemann problem at the interface between two cells in a numerical
scheme inside of the IACS. The 5-wave Riemann fan is illustrated
(the slow waves are ignored without loss of generality) in order to
show the difference of the resolved flux at the interface calculated
with a 2-wave HLL Riemann solver and the same calculation performed
with a 5-wave HLLI Riemann solver. Outgoing and ingoing are defined
in the frame of reference of the plasma.

\item HLL Riemann Solver vs. HLLI Riemann Solver  Stationary Alfven Discontinuity. Figure 2 shows two simulations of an isolated, stationary
contact discontinuity. The density variable is shown. The result
from the HLL Riemann solver is shown with crosses while the result
from the HLLI Riemann solver is shown with dots.

\item HLL Riemann Solver vs. HLLI Riemann Solver Contact Discontinuity. Figure 3 shows two simulations of an isolated, stationary
Alfven wave. The transverse velocity and magnetic field are shown.
The result from the HLL Riemann solver is shown with crosses while
the result from the HLLI Riemann solver is shown with dots.

\item Torsional Alfven Wave
Dissipation Test, Figure 4 shows the results of the torsional Alfven
wave dissipation test. A second order WENO reconstruction was used
in all these tests. Figure 4a shows the decay in the z-component of
the velocity as a function of time. Figure 4b shows the same for the
z-component of the magnetic field.

\item MC Limiter. Figure 5 is analogous to Figure 4 with the
exception than an MC limiter was used. The MC limiter is considered
inferior to a good WENO scheme. Comparing Figures 4 and 5, this
observation is apparent in the figures.

\item ADER-WENO Schemes the 2D MuSIC Riemann Solver. Figure 6 shows the same Alfven wave propagation
test. This time, we used different ADER-WENO schemes with increasing
order of accuracy. We also used the 1D HLLI Riemann solver along
with the 2D MuSIC Riemann solver with sub-structure. We see that
higher order schemes produce lower dissipation.

\item ADER-WENO Schemes and the 2D HLL Riemann Solver. Figure 7 shows the same Alfven wave propagation test as Figure 6 but when a lower quality Riemann solver is used.
We used the same ADER-WENO schemes as in Figure 6. The only
difference is that the simulations in this Figure were run with a 1D
HLL Riemann solver and a 2D HLL Riemann solver. Comparing Figure 6
to Figure 7 enables us to appreciate the improvement that Riemann
solvers with sub-structure provide in reducing dissipation.

\end{enumerate}

\end{document}